\documentclass[12pt]{article}
\usepackage{hyperref}
\usepackage{cite}
\usepackage{color}
\usepackage{graphicx}
\usepackage{amsmath}
\usepackage{amssymb}
\usepackage{xspace}

\makeatletter
\@addtoreset{equation}{section}

\makeatletter
\renewcommand\section{\@startsection {section}{1}{\z@}%
                                   {-3.5ex \@plus -1ex \@minus -.2ex}
                                   {2.3ex \@plus.2ex}%
                                   {\normalfont\large\bfseries}}
\renewcommand\subsection{\@startsection{subsection}{2}{\z@}%
                                     {-3.25ex\@plus -1ex \@minus -.2ex}%
                                     {1.5ex \@plus .2ex}%
                                     {\normalfont\bfseries}}

\def\baselinestretch{1.2}
\newcommand{\be}{\begin{equation}}
\newcommand{\ee}{\end{equation}}
\newcommand{\beq}{\begin{eqnarray}}
\newcommand{\eeq}{\end{eqnarray}}

\newcommand{\gone}[1]{{}}

\begin{document}
\begin{titlepage}
\begin{flushright}
MAD-TH-14-07
\end{flushright}

\vfil

\begin{center}

{\bf \Large
Solitons  on Intersecting 3-Branes\\ II: a Holographic Perspective}

\vfil

William Cottrell, Akikazu Hashimoto, Duncan Pettengill, and Mohandas Pillai

\vfil

Department of Physics, University of Wisconsin, Madison, WI 53706, USA

\vfil

\end{center}

\begin{abstract}
\noindent We study the low energy effective theory of two sets of
D3-branes overlapping in 1+1 dimensions, recently considered by
Mintun, Polchinski, and Sun. In the original treatment by MPS, by studying the
properties of magnetic solitons, the low
energy effective field theory was found to require some ultraviolet
completion, possibly involving full string dynamics.  Recently in a companion paper, it
was shown that by scaling the angle between the D3-branes and the
D3'-branes in the zero slope limit in specific way, one can find
simpler effective field theory which consists of a single tower of
Regge trajectory states and yet is ultraviolet complete and
non-singular. In this article, we study this model by further studying
a limit which recovers the MPS dynamics from this non-singular
construction. We approach this issue from a holographic perspective,
where we consider a stack of $N$ D3-branes overlapping with a single
D3'-brane, and treat that D3'-brane as a probe in the $AdS_5 \times
S^5$ dual. In general, the D3'-brane probe supports a magnetic
monopole as a non-singular soliton configuration, but in the limit
where the MPS dynamics is recovered, the soliton degenerates. This is
consistent with the idea that the effective dynamics in the MPS
setup is incomplete, but that it can be completed with a single tower
of Regge trajectory states.

\end{abstract}
\vspace{0.5in}

\end{titlepage}
\renewcommand{\baselinestretch}{1.05}  

\section{Introduction}

Recently, Mintun, Polchinski, and Sun studied a simple intersecting D-brane system consisting of a D3-brane intersecting a D3'-brane  \cite{Mintun:2014aka}. The 3-branes were oriented as  follows:
\be
\begin{tabular}{c||cccccccccc}
       & 0& 1  & 2& 3& 4& 5& 6& 7& 8& 9 \\
       \hline
D3 & $\circ$ &  $\circ$ & $\circ$ &  $\circ$ &    &   &    &   &   &     \\
D3' & $\circ$ &   && $\circ$ & $\circ$ & $\circ$ &    &   &    \\
\end{tabular} \label{orientation}
\ee
In other words, they are overlapping in 1+1 dimensions along the $x_0$
and $x_3$ directions. The D3-branes were also arranged to be separated
by a finite distance of order $\alpha' V$ along the $x_6$
direction. The low energy spectrum arising from the open strings in
such a setup is easy to infer. We expect to find ${\cal N}=4$ $U(1)$
Yang-Mills theory on the world volume of the D3, and another ${\cal
  N}=4$ $U(1)$ Yang-Mills theory on the world volume of the D3'. In
addition, one expects to find a ${\cal N}=2$ $d=3+1$ hypermultiplet
fields $B$ and $C$ charged as a bifundamental in $U(1) \times U(1)$
with mass $m^2 = V^2$, dimensionally reduced to 1+1 dimensions. These
fields are then coupled to the $U(1)$ fields along a defect in such a
way to preserve a total of eight supercharges
\cite{Constable:2002xt}. One is more or less led to a unique low
energy effective action following this procedure
\cite{Mintun:2014aka}.

A natural question considered by \cite{Mintun:2014aka} is whether such
an action defines a complete dynamical system as a quantum theory,
arising as a systematic $\alpha'\rightarrow 0$ limit of the brane
construction outlined above.  One diagnostic for this issue is whether
the magnetic duals of the $B$ and $C$ fields, which should arise as a
D-string stretching between the D3 and the D3' branes, would exist as
a BPS soliton of the candidate decoupled theory. The existence of such
a soliton would be expected since the brane configuration which we
started from is manifestly S-duality invariant, and we expect that
property to survive the $\alpha' \rightarrow 0$ limit since both the
fundamental string and the D-string stretching between the D3 and the
D3' have finite mass in the scaling limit.

It therefore came as somewhat of a surprise when the conclusion of
\cite{Mintun:2014aka} to this question was {\it negative}. The naive
candidate Lagrangian failed to support a soliton with the required
property. With some effort, \cite{Mintun:2014aka} proposed a
modification to the candidate effective theory so that a soliton can
be supported. This involved generalizing the metric on the field space
of $B$ and $C$ fields to a broader K\"ahler class. While
\cite{Mintun:2014aka} reported some success with this approach, their
ultimate conclusion was that they are unable to avoid a singularity in
their metric, signaling that some ultraviolet completion is required
in order to fully regulate the dynamics.

More recently a simple generalization of the intersecting D3 system
was studied \cite{Cottrell:2014ura}.  The generalization consisted of
slight change in the scaling of the angle between the D3 and the D3'
brane. Instead of configuring the branes to be perpendicular in the 14
and 25 planes, one can set the orientation as follows:\\
\be \begin{tabular}{ll}
D3: & 0 $( {1 \atop 4})_{-\gamma}({2 \atop 5})_{\gamma}$   3\\
D3': & 0 $( {1 \atop 4})_{\gamma}({2 \atop 5})_{-\gamma}$   3
\end{tabular} \label{orient1}
\ee
and scale $\gamma$ so that
\be \tan \gamma = \alpha' a \label{scaling}\ee
where $a$ is a parameter with dimension of mass squared.

Schematically, the brane intersection of \cite{Cottrell:2014ura} is
identical to that of \cite{Mintun:2014aka}. In both constructions, the
brane overlap along 1+1 dimensions. The main difference, however,
stems from the spectrum of the 33' states. When the angle is scaled according to (\ref{scaling}), the 33' spectrum will consist of a tower of states with mass
\be m^2  \sim V^2 + a n, \qquad n = 0, 1, 2, \ldots  \label{tower} \ee
which remains finite in the $\alpha' \rightarrow 0$ limit. This basic
feature was observed originally in \cite{Hashimoto:1997gm}. With the
angle scaled according to (\ref{scaling}), the system can be
compactified and T-dualized to an ordinary $SU(2)$ gauge theory with a
non-vanishing non-abelian flux. The spectrum of small fluctuations
around this background was worked out by van Baal in
\cite{vanBaal:1984ar}. The tower states is essentially the
Landau-level in response to the constant non-abelian magnetic
field. In the limit $a \rightarrow 0$, these states become momentum
modes of the off-diagonal components of the $SU(2)$ gauge fields.

In other words, the scaling (\ref{scaling}) gives rise to a more
conventional field theory description of the intersecting brane
configuration compared to the case when the branes are arranged to be
perpendicular as was done in \cite{Mintun:2014aka}.  It is natural
then to consider if the magnetically charged solitons that
\cite{Mintun:2014aka} sought exists in the effective theory with a
tower of states (\ref{tower}). That was the question which was
addressed in \cite{Cottrell:2014ura}. It should not come as a big
surprise that the answer to this question is {\it positive}, i.e. a
magnetic monopole soliton does exist for this system.  An explicit
form of the soliton solution is not known as of yet.  Nonetheless, an
explicit existence proof was presented in \cite{Cottrell:2014ura}.

Assuming that the soliton exists and is perfectly well behaved in the
scaling (\ref{scaling}), the interesting issue to contemplate is what
happens when we take the limit $a \rightarrow \infty$. In this limit,
all but the $n=0$ state in the tower (\ref{tower}) decouples. The
$U(1)\times U(1)$ degrees of freedom also survives. In other words, we
recover the spectrum of states originally considered by
\cite{Mintun:2014aka}. So, if the conclusion that the soliton is
absent in \cite{Mintun:2014aka} is correct, the soliton found in the
scaling (\ref{scaling}) should somehow degenerate in the $a
\rightarrow \infty$ limit. Unfortunately, without the explicit form of
the soliton solution, it is difficult to study if and how this is
happening.

In this article, we will probe this issue from a holographic
perspective. The idea is to consider a stack of $N$ D3-branes, which
we describe as a gravity background, and treat the D3' as a probe. In
other words, we will orient our branes as follows:\\
\be \begin{tabular}{rl}
$N$ D3\,:  & 0123 \\
D3': & 0 $( {1 \atop 4})_{\gamma}({2 \atop 5})_{-\gamma}$   3
\end{tabular} \label{orient2}
\ee
Then, we will scale\footnote{The $c$ here is related to the $a$ of \cite{Cottrell:2014ura} via $\tan^{-1}(R^{2}c)=2\tan^{-1}(2\pi\alpha' a)$. Some factors of $\lambda$ enter in these relations for notational convenience.}
\be \tan \gamma = R^2 c = \sqrt{\lambda} \alpha' c \label{cscale} \ee
and take the de-coupling limit $\alpha' \rightarrow 0$. At this point,
our problem becomes that of embedding a D3'-brane in $AdS_5 \times
S^5$. We will be interested in a particular embedding where there will
be a unit of magnetic charge on the D3' world volume. In the following
sections, we will outline the steps needed to study such an
embedding. We will then conclude by describing how the soliton behaves
as the limit $c \rightarrow \infty$ is taken.

\section{D3'-brane embedding}
\label{geometry}

In this section, we will review the basic setup for describing the embedding of a D3'-brane probe in $AdS_5 \times S^5$ geometry. 

\subsection{Supergravity background}

Let us begin by reviewing the $AdS_5\times S^5$ geometry to setup our notations and conventions. The background geometry and flux is given by
\begin{eqnarray}
ds^{2}&=&R^{2}\left(u^{2}\eta_{\mu\nu}dx^{\mu}dx^{\nu}+\frac{du^{2}}{u^{2}}+d\Omega_{5}^{2}\right) \\ \nonumber
F_{5} &=& 4R^{4} \left(1+*\right) d\Omega_{5}
\end{eqnarray}
$R$ is the $AdS$ radius
\begin{equation}
R^{4}=4\pi g_{s}N(\alpha')^{2} \equiv \lambda \alpha'^{2}
\end{equation}
and
\be \lambda = 4 \pi g_s N = 2 g_{YM}^2 N \ee
is the 't Hooft coupling from the $SU(N)$ field theory perspective.  In order for the semi-classical treatment of the D3'-brane probe to be effective, we take $\lambda$ to be {\it large} but {\it finite}. 

\subsection{DBI action for the D3'-brane probe}

The D3'-brane probe will be arranged to be extended along the 0123
directions, and embedded non-trivially in the 456789 directions. We
can therefore use $x^\mu$ with $\mu=0,1,2,3$ as the world volume
coordinate for the D3'-brane probe. Because of the symmetry, it will
also turn out to be convenient to use cylindrical world volume
coordinates $(t,\rho,\varphi,x_3)$.

We will parameterize the transverse coordinates
  coordinates $x^{4}\ldots
x^{9}$ into which the D3'-brane is embedded in polar coordinates as follows:
\begin{eqnarray}
x^{6}&=&r \cos\theta \\
x^{4}&=& r\sin\theta \cos\phi \nonumber \\
x^{5} &=& r\sin\theta \sin\phi\cos\alpha_{1} \nonumber \\
x^{7}&=&r\sin\theta\sin\phi\sin\alpha_{1}\cos\alpha_{2} \nonumber \\
x^{8}&=&r\sin\theta\sin\phi\sin\alpha_{1}\sin\alpha_{2}\cos\alpha_{3}\nonumber  \\ 
x^{9}&=&r\sin\theta\sin\phi\sin\alpha_{1}\sin\alpha_{2}\sin\alpha_{3} \ .  \nonumber
\end{eqnarray}
For our purposes it will be sufficient to truncate to $x^{7,8,9}=0$. In other words, we restrict our attention to the case where $\alpha_1=\alpha_2=\alpha_3=0$.
Thus, the dynamical variables are $x^{4,5,6}$, or, equivalently, $r$,
$\theta$, and $\phi$. 

In taking the near horizon limit, we will scale
\begin{equation}
\label{rurel}
r=R^{2}u
\end{equation}
and keep $u$ fixed as $\alpha' \rightarrow 0$.\footnote{Note that this differs
by a factor of $\sqrt{\lambda}$ from the scaling convention where
$U=r/\alpha'$ is kept fixed, e.g., in \cite{Maldacena:1997re}. Since
$\lambda$ is kept large but finite, this is strictly speaking the same
scaling, though some care is necessary in keeping track of quantities
being kept fixed when analyzing large $\lambda$ asymptotics.}

 When treating the transverse scalars as fields
via the AdS/CFT correspondence we will adopt the notation:
\begin{equation}
\Phi_{i}=\frac{x^{i}}{R^{2}}=\frac{x^{i}}{\sqrt{\lambda}\alpha'}, \qquad (i=4,5,6) \ . 
\end{equation}
A static D3'-brane embedding is now parameterized by $u(\rho,\varphi,x_3)$, $\theta(\rho,\varphi,x_3)$, and $\phi(\rho,\varphi,x_3)$. The cylindrical symmetry immediately allows one to solve
\be \phi = \varphi \ee
and treat $u(\rho,x_3)$ and $\theta(\rho,x_3)$ as being independent of
$\varphi$. Our task now is to find the equation of motion for the
static embedding.

For this, we consider the DBI action. By explicitly computing the pullback metric and the 4-form potential, we find
\begin{eqnarray}
\label{action}
I_{D3}&=&I_{DBI}+I_{WZ} \\ 
I_{DBI} &=& -T_{3}\int d^{4}x\,  e^{-\Phi}\sqrt{-\det\left(g_{ij}+\mathcal{F}_{ij}\right)} \\ \nonumber
&=&-T_{3}\int d^{4}x \Bigg{(}u^{4}\left(4\pi^{2}(F_{\varphi 3}^{2}+F_{\rho\varphi})^{2}+u^{4}\rho^{2}+u^{2}\sin^{2}\theta\right) \\ \nonumber
&&+4\pi^{2}\left((\partial_{3}u)^{2}\left(F_{\varphi 3}+F_{\rho \varphi}\right)^{2}+\left(\partial_{\rho}\theta F_{\varphi 3}+\partial_{3}F_{\rho \varphi}\right)^{2}u^{2}\right) \\ \nonumber
&&+u^{4}\left(2(\partial_{3}u)^{2}+\left((\partial_{3}\theta)^{2}+(\partial_{\rho}\theta)^{2}\right)\right)\rho^{2}\\ \nonumber
&&+u^{2}\left(2(\partial_{3}u)^{2} +\left((\partial_{3}\theta)^{2}+(\partial_{\rho}\theta)^{2}\right)u^{2}\right)\sin^{2}\theta \\ \nonumber
&&-\frac{1}{2}(\partial_{3}u)^{2}(\partial_{3}\theta-\partial_{\rho}\theta)^{2}\left(-1+2u^{2}\rho^{2}+\cos2\theta\right)\Bigg{)}^{1/2}\\
I_{WZ}&=& T_{3} \int e^{\mathcal{F}} \wedge C \\ 
&=&T_{3} \int d^{4}x\,  R^{4}u^{4}\ , \nonumber 
\end{eqnarray}
where $\mathcal{F}=B+2\pi \alpha' F$, $F$ is the world volume field
strength, and $B=0$ in our background.

All that remains to be done, then, is to analyze the equation of motion for the embedding fields $u(\rho,x_3)$ and $\theta(\rho,x_3)$.

\subsection{Constraints due to supersymmetry}
\label{supersymmetry}

The action as written in (\ref{action}) gives rise to a rather
formidable set of equations of motion. It would be a prohibitive task
to analyze our problem that way. Fortunately, the static configuration
we seek is expected to preserve four supercharges. Generally,
Born-Infeld action restricted to supersymmetric configurations exhibit
dramatically simpler behavior \cite{Callan:1997kz}. Indeed, 
following the analysis of $\kappa$-symmetry for D3-branes embedded in
$AdS_5\times S^5$ originally carried out in \cite{Skenderis:2002vf}, we infer\footnote{See Appendix \ref{appendixA} for details}.
\begin{eqnarray}
\label{finalansatz}
2\pi\alpha' F_{\varphi3}&=&R^{2}\left(u\sin^{2}\theta-\rho\partial_{\rho}u\right)\sec\theta \\ \nonumber
2\pi\alpha' F_{\rho\varphi}&=&-\frac{R^{2}}{u^{2}\rho}\left(u^{2}\rho^{2}+\sin^{2}\theta\right)\partial_{3}u \sec\theta \\ \nonumber
\partial_{\rho}\theta&=&\frac{u-\rho\partial_{\rho}u}{u\rho}\tan\theta \\ \nonumber
\partial_{3}\theta&=&-\frac{1}{u}\tan\theta\partial_{3}u \ . 
\end{eqnarray}
The last two equations  (\ref{finalansatz}) can be integrated to read
\begin{equation}
\label{thetacond}
u\sin\theta=c \rho
\end{equation}
where $c$ is an integration constant.  Recalling that we have already constrained $\phi = \varphi$, we find that
\be \Phi_4+ i \Phi_5  = u \sin\theta e^{i \varphi} = c \rho e^{i \varphi}\ee
we see that the integration constant $c$ that appear here is precisely
the same as the one parameterizing the scaling (\ref{cscale}).

The remaining constraint from the first two equations   in (\ref{finalansatz})  takes a simple form when parameterized in terms of 
\begin{equation}
\label{phiconvention}
\Phi_{6}=\frac{x^{6}}{R^{2}} = u\cos\theta \ . 
\end{equation}
They take the form
\begin{eqnarray}
\label{finans}
2\pi \alpha' F_{\varphi3}&=&-R^{2}\rho\partial_{\rho}\Phi_{6} \\ \nonumber
2\pi\alpha' F_{\rho \varphi}&=&R^{2}\rho\left(1+\frac{c^{2}}{\left((c\rho)^{2}+\Phi_{6}^{2}\right)^{2}}\right)\partial_{3}\Phi_{6} \ . 
\end{eqnarray}

Since the $F$'s appearing on the left hand side of (\ref{finans}) are a $U(1)$ field strength, they must satisfy the Bianchi identity, which constrains $\Phi_6$ to satisfy 
\begin{equation}
\label{eom}
{1 \over \rho} \partial_\rho \left(\rho \partial_\rho \Phi_{6}\right)
+\partial_{3}\left(1+\frac{c^{2}}{\left((c\rho)^{2}+\Phi_{6}^{2}\right)^{2}}\right)\partial_{3}\Phi_{6}=0 \ . 
\end{equation}

This is a {\it second order}, {\it non-linear}, partial differential
equation governing the embedding of D3'-brane in $AdS^5$. This is the
main equation which we will refer to as the {\it full embedding
  equation}. However, since the analysis leading up to the derivation of this
equation (\ref{eom}) was somewhat involved, it would be
useful to subject it to some simple tests. We will perform a few such
tests in the remainder of this section, and continue with the analysis
of (\ref{eom}) in the next section.

\subsection{Tests of the full embedding equation}

\subsubsection{BPS Energy formula}

One feature of the supersymmetric configuration of the Born-Infeld
system is that the argument of the square root becomes a perfect
square, making the action rational. This is indeed the case. If we substitute the constraints  (\ref{thetacond}) and (\ref{finans}) into the full action (\ref{action}), we find
\begin{eqnarray}
\label{bpsaction}
\mathcal{L}=-\frac{\lambda}{8\pi^{3}g_{s}}\left(\rho c^{2}+\rho\Bigg{(}(\partial_{3}\Phi_{6})^{2}+(\partial_{\rho}\Phi_{6})^{2}+c^{2}\left(\frac{(\partial_{3}\Phi_{6})^{2}}{((c\rho)^{2}+\Phi_{6}^{2})^{2}}\right)\Bigg{)}\right) \ . 
\end{eqnarray}
As expected, the final expression does not involve any square
roots. It should be stressed, however, that variation of
(\ref{bpsaction}) with respect to $\Phi_6$ will {\it not} give rise to
(\ref{eom}). The reason is that $\Phi_6$ was constrained through
(\ref{finans}) and can not varied as if it were an unconstrained
field. Applying suitable Lagrange multipliers to respect the
constraint will give rise to (\ref{eom}).

The utility of (\ref{bpsaction}) rests in the fact that it provides
the measure of energy density. The term $\rho c^{2}$ gives rise to a
uniform energy density that can be attributed to the tension of an
ordinary tilted brane. Subtracting this divergent piece would leave
\begin{equation}
\label{energy}
E=\frac{\lambda}{8\pi^{3}g_{s}}\int d\rho d\varphi dx^{3} \, \rho  \left((\partial_{3}\Phi_{6})^{2}+(\partial_{\rho}\Phi_{6})^{2}+c^{2}\left(\frac{(\partial_{3}\Phi_{6})^{2}}{((c\rho)^{2}+\Phi_{6}^{2})^{2}}\right)\right)
\end{equation}
which can be used to compute the energy of a  soliton solving the full embedding equation (\ref{eom}).

\subsubsection{Tilted Brane\label{sec:tiltedbrane}}

One simple solution to (\ref{eom}) is the plane tilted brane
embedding.  The solution is 
\be \Phi_6=v 
\ee
some constant $v$ parameterizing the distance $\Delta x_6 = R^2 v$ of
the D3' probe from the horizon at its point of closest separation, but
with the understanding that the constraint (\ref{thetacond}) is
applied with non-vanishing $c$ so that the brane is tilted relative to
the horizon. The energy (\ref{energy}) for this configuration is
identically zero.

\subsubsection{Magnetic monopole on untilted D3'-brane}

The final diagnostic example we will consider is to set $c=0$, so that
the D3'-probe is interpreted as describing the $U(1)$ component of
the dynamics in the Coulomb branch $SU(N+1) \rightarrow SU(N) \times U(1)$

The simplicity of the $c=0$ case is obvious from looking at the full embedding equation (\ref{eom}), which reduces to the Laplace equation
\begin{equation}
\label{laplace}
{1 \over \rho} \partial_\rho \left(\rho \partial_\rho \Phi_{6}\right)
+
\partial_3^2 \Phi_{6}
=0
\end{equation}
which is solved by 
\begin{equation}
\label{monsol}
\Phi_{6}=v-\frac{q}{r'}
\end{equation}
where $r'=\sqrt{\rho^{2}+(x^{3})^{2}}$.  Precisely this solution was discussed in a recent paper by Schwarz \cite{Schwarz:2014rxa}. Here, $v$ describes the asymptotic value of $\Phi_6(\rho,x_3)$ away from the monopole.

The value of $q$ is determined by charge quantization.   Using (\ref{finans}), we have
\be
\label{fmonopole}
\int F_{\theta'\varphi}
={2 \alpha'} qR^{2} = 2 q  \sqrt{\lambda} = 2 \pi k  
\ee
where $k$ must take on integer values because of the Dirac quantization condition. So,
\begin{equation}
\label{quantcond}
q=\frac{\pi k}{\sqrt{\lambda}} \ . 
\end{equation}
We will set $k=1$ to describe a singly charged monopole. 

\begin{figure}\center{
\includegraphics[scale=0.8]{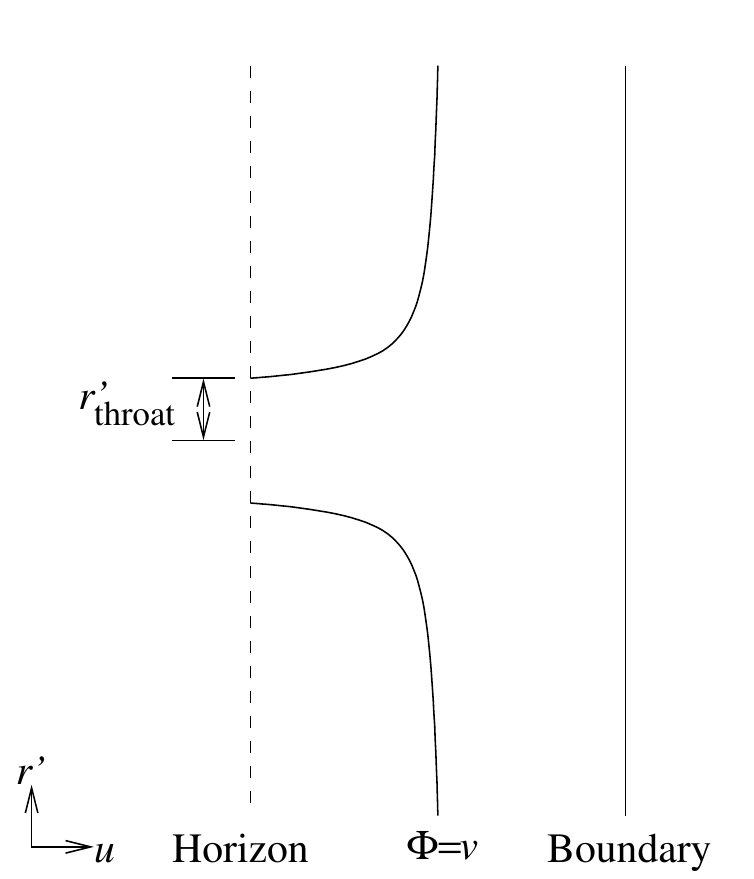}}
\caption{Schematic form of the D3'-brane embedding corresponding to a
  magnetic monopole soliton in ${\cal N}=4$ SYM in Coulomb branch
  $U(N+1) \rightarrow U(N) \times U(1)$. The $U(1)$ component is
  manifested as a D3'-probe embedded in $AdS_5\times S^5$ in Poincare
  patch.  The vertical axis is $r'$ and the horizontal axes is the
  radial coordinate $u$. The D3'-probe melts into the horizon at
  $u=0$.  \label{simpleMonopole}}
\end{figure}

The solution is cut-off at $\Phi_6=0$ where the D3' hits the horizon. This happens at 
\begin{equation}
r'_{throat}=\frac{\pi}{\sqrt{\lambda} v} \ . 
\end{equation}
The schematic form of this embedding is illustrated in figure \ref{simpleMonopole}.  Note that even though the throat has finite coordinate size $r'_{throat}$, its geodesic size is zero. 

The energy of this monopole can be computed from (\ref{energy}) for $c=0$ which reads
\be
E_{mon}
= \frac{\lambda}{(2\pi)^{3}g_s}\int^{\infty}_{r_{throat}} dr' 4\pi r'^{2} (\partial_{r'}\Phi_{6})^{2} 
=\frac{\lambda}{2\pi^{2}g_{s}}\int^{v}_{0}q d\Phi_{6}
=\frac{1}{2\pi g_{s}} \sqrt{\lambda} v   \label{Emon} 
\ee
where the form of the solution (\ref{monsol}) was used in one of the steps. 
The  energy is precisely that of a D1-string stretched over a distance
\be \Delta x_6= R^2 v \ . \ee
Using $g^{2}_{YM}=2\pi g_{s}$ this may finally be written as:
\begin{equation}
E_{mon}=\frac{1}{g^{2}_{YM}} \sqrt{\lambda} v \ . 
\end{equation}
A factor of $\lambda$ may seem unfamiliar, but that is because we defined 
\be R^2 v =\alpha' V \ee
and so in terms of $V$, we recover the more familiar looking expression
\begin{equation}
E_{mon}=\frac{1}{g^{2}_{YM}} V
\end{equation}
that one finds, for instance, in \cite{Weinberg:2006rq}.

\section{Magnetic monopole soliton solution on a tilted brane}

In this section, we describe solutions to the full embedding equation
(\ref{eom}) corresponding the magnetic monopole soliton on a tilted
D3'-brane (with non-vanishing $c$) in $AdS_5 \times
S^5$. Unfortunately, the full non-linear form of (\ref{eom}) is rather
formidable to analyze in closed form. Fortunately, most of the
interesting features can be extracted from a linearized approximation
where we expand $\Phi_6$ around its asymptotic background value
$v$. In the following we will summarize this approximation
and describe the scope of its validity.

\subsection{Linearization}

If we substitute
\begin{equation}
\Phi_{6}=v+\delta \Phi_{6}
\end{equation}
into (\ref{eom}) and only keep the terms linear in $\delta \Phi_6$, we obtain an equation
\begin{equation}
\label{philin}
{1 \over \rho} \partial_\rho \left(\rho \partial_\rho \delta \Phi_{6}\right)
+\left(1+\frac{c^{2}}{\left((c\rho)^{2}+v^{2}\right)^{2}}\right)\partial_{3}^2\delta \Phi_{6}=0
\end{equation}
which is much more manageable than (\ref{eom}). This equation can be understood as a Laplace equation for the metric
\begin{equation}
\label{rescaledmetric}
ds^{2}=-dt^{2}+(dx^{3})^{2}+ \left(1 + \frac{c^{2}}{(v^{2}+(c\rho)^{2})^{2}}\right)\left(d\rho^{2}+\rho^{2}d\phi^{2}\right)
\end{equation}
which is conformally equivalent to the pull-back of the flat tilted brane described in Section \ref{sec:tiltedbrane}. 

From simply examining the form of the metric (\ref{rescaledmetric}),
we see that there are features at scales $\rho = v/c$ and $\rho =
1/\sqrt{c}$. Let us take ${v^2/ c} \ll 1$ to keep these scales
separated parametrically. Then, our geometry (\ref{rescaledmetric})
can be divided into three distinct regions where the metric simplifies locally.

\begin{itemize}
\item {\bf Region I:} $\rho \gg 1/\sqrt{c}$

This is the large radius asymptotic region. In this limit, the metric
(\ref{rescaledmetric}) is flat.
\begin{equation}
ds^{2}= -dt^{2}+(dx^{3})^{2}+d\rho^{2}+\rho^{2}d\phi^{2} \ . 
\end{equation}

\item {\bf Region II:} $v/c \ll \rho \ll 1/\sqrt{c}$

In this region the metric is approximately:

\begin{equation}
ds^{2}=-dt^{2}+(dx^{3})^{2}+\frac{1}{c^{2}\rho^{4}}\left(d\rho^{2}+\rho^{2}d\phi^{2}\right) \ . 
\end{equation}

Under the change of variables $\rho = 1/(c y)$ this becomes flat:

\begin{equation}
ds^{2}=-dt^{2}+(dx^{3})^{2}+y^{2}d\phi^{2}+dy^{2} \ . 
\end{equation}

\item {\bf Region III}:  $\rho \ll v/c$

Finally, in this small $\rho$ region,  the metric is  once again flat but the $(\rho,\phi)$ plane is rescaled.
\begin{equation}
ds^{2}=-dt^{2}+(dx^{3})^{2}+\left(\frac{c^{2}}{v^{4}}\right)\left(d\rho^{2}+\rho^{2}d\phi^{2}\right) \ . 
\end{equation}
\end{itemize}

In order for all of these regions to exist, we must set $v^2/c \ll
1$. This is the limit that is interesting when taking the large $c$ limit. In the opposite, small $c$ limit, Region II disappears and the metric in region III becomes instead 
\begin{equation}
ds^{2}=-dt^{2}+(dx^{3})^{2}+\left(1+\frac{c^{2}}{v^{4}}\right)\left(d\rho^{2}+\rho^{2}d\phi^{2}\right)
\end{equation}
which is essentially the same as the flat space metric in Region I
with a minor rescaling in the $(\rho,\phi)$ plane.

\subsection{General features of the linearized solution}

At this point, it is rather straightforward to argue that a solution
to the Laplace equation with a point-like source on metric
(\ref{rescaledmetric}) will generically exist. The background metric
is locally smooth and asymptotically flat. The only remaining issue is
how this solution behaves in the limit that $c$ is taken to be large,
keeping $v$ and $\lambda$ fixed. In order to address this issue, let
us further explore the behavior of the solution to (\ref{philin}) with
a suitably normalized point-like source at the origin for small
$v^2/c$. 

One way to proceed is to separate variables 
\be \delta \Phi_6(x_3,\rho) = \int {dk \over 2 \pi} \,c_k e^{i k x_3} \psi_k(\rho) \ee
and then re-combine $\psi_k$ along the lines of
\cite{Cherkis:2002ir}. We will perform some part of this analysis in
the next section, but it turns out to be a rather cumbersome exercise
in light of the fact that the equation for $\psi_k(\rho)$ is still
rather complicated.

Fortunately, one can gain some intuition by studying the behavior of geodesic distance from the origin where a localized source is placed. On flat space, 
\be \delta \Phi = v - {q \over d} \ee
where $d$ is the geodesic distance, is the correct exact
solution. Analyzing the same quantity for the metric
(\ref{rescaledmetric}) and subjecting it to some tests can provide
quite a bit of intuition on the behavior of the solution we are trying
to study.

\begin{figure}
\centerline{\includegraphics[scale=0.8]{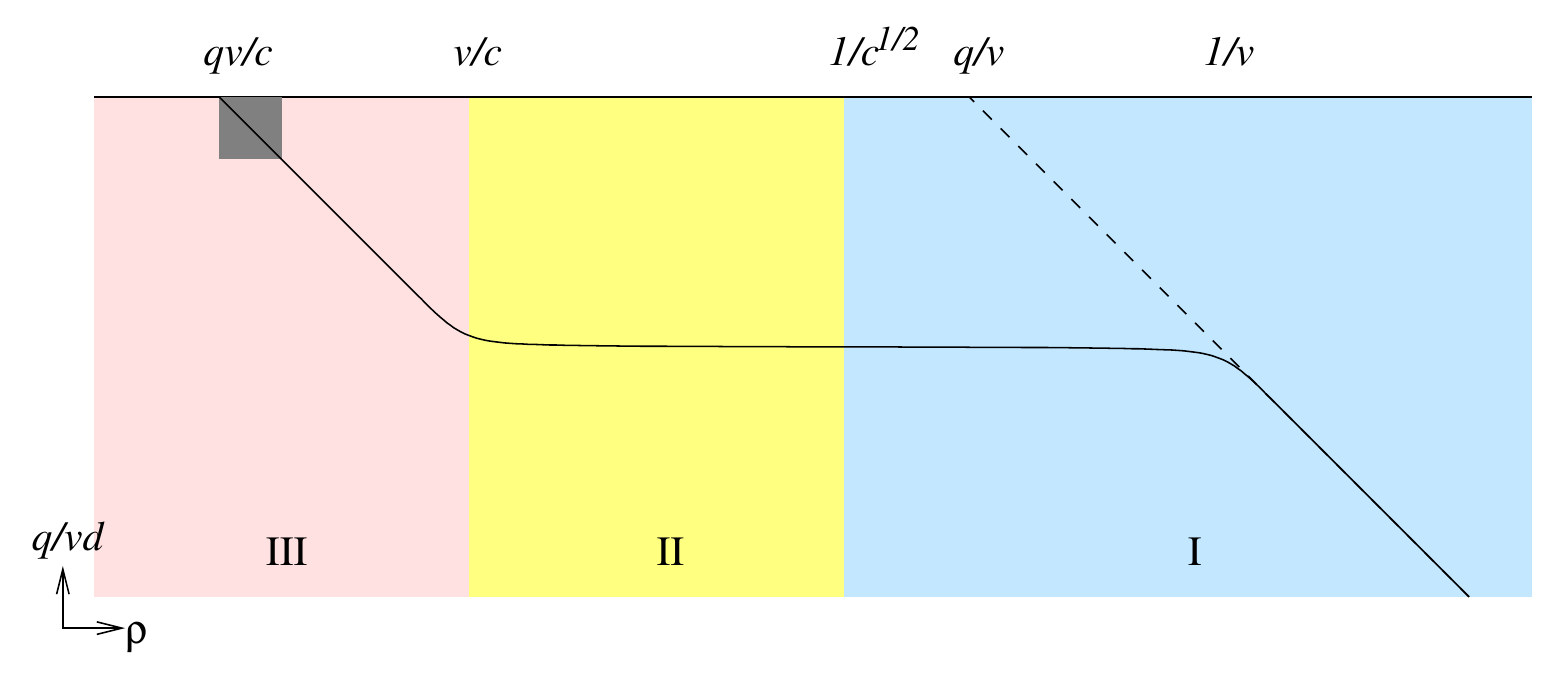}}
\caption{Logarithmic plot of $q/vd$ v.s. $\rho$. For this plot,
  $q=\pi/\sqrt{\lambda} = 10^{-4}$ and $v^2/c = 10^{-12}$. Regions I,
  II, and III are indicated with using different background colors.
  The point where $q/vd = 1$ is where the $\Phi_6=0$. The shaded box
  is the region where the corrections to the linearized approximation
  are expected to be important. The dotted line is the untilted
  solution (\ref{monsol}). \label{figb}}
\end{figure}

Let us consider the geodesic as a function of $\rho$ where $x_3$ is fixed to zero. Then,
\be d(\rho) = 
\int_0^{\rho} d \rho'\, \sqrt{1 + \frac{c^{2}}{(v^{2}+(c\rho')^{2})^{2}}} \ . 
\ee
It is not too difficult to compute this numerically for some fixed $c$
and $v$. In order to illustrate all the hierarchically separated scales, it is convenient to display the potential in a log-log plot, of 
\be \log\left(1 - {\delta \Phi \over v}\right) = \log \left({q \over v d}\right) \ee
as a function of
\be \log\left({c \rho \over v}\right) \ . \ee
The result of such analysis is illustrated in  figure \ref{figb}.

Several features are notable in the plot illustrated in figure
\ref{figb}. First of all, $q/vd$ exhibits a $\rho^{-1}$ scaling
behavior in region III, but becomes approximately flat in region
II. That approximatly flat behavior continues into region I, but then
the curve bends and asymptotes again into a homogeneous $\rho^{-1}$
scaling behavior. All of these features are consistent with a more
careful analysis which we will describe in greater detail in the next
section. This bending in region I can be viewed, from the perspective
of observers at large $\rho$, as the charge being effectively smeared
along the $x_3$ direction. The extent that this is happening will be
further discussed in the next section.

This plot also highlights how one should think about the limit $c
\rightarrow \infty$ keeping $v$ and $q = \pi/\sqrt{\lambda}$
fixed. The plot simply gets wider in the horizontal direction. On the
other hand, when $\lambda$ is increased keeping $v$ and $c$ fixed, the
entire plot in figure \ref{figb} simply slides downward. The features
illustrated in figure \ref{figb} become more reliable when $\lambda$
is taken to be large.

\subsection{Soliton Energy\label{energetics}}

Let us now consider the energy contained in the monopole solution. The
solution to the Laplace equation (which we have not found in explicit
form) will have a definite energy density when substituted back into 
(\ref{energy}) in the linearized approximation.  A convenient way
to parameterize the D3'-brane probe world volume is in terms of
contours of fixed $\delta \Phi_6$. Then, just as was the case for the
untilted monopole (\ref{Emon}), the energy can be computed as
\be E = {\sqrt{\lambda} \over 2 \pi g_s} \Delta \Phi_6\label{DPHI} \ee
so as long as the fixed $\Phi$ contour from $\Phi=0$ and $\Phi=v$
covers the region of the D3'-probe outside the horizon, we would get
the expected BPS energy for the solitons. The statement (\ref{DPHI})
can also be generalized to the fully non-linear case for the soliton
solving (\ref{eom}) and applied to the energy formula (\ref{energy}).

\subsection{Validity of the linearized approximation}

There is one subtlety we must confront in the linearized treatment of
the embedding equation (\ref{eom}).  Unlike in the untilted case where
the entire embedded brane coincided with the horizon when $\Phi_6=0$,
for the tilted brane,in the configuration illustrated in figure
\ref{figb}, this is not exactly the case. This is because at
$\rho\approx qv/c$, the embedded D3' is separated from the horizon by
distance $R^2 c \rho \approx \alpha' v$. Strictly speaking, this is a
substringy scale, but so is everything separated in the typical
radial positions in $AdS_5 \times S^5$.

There is, however, a way out of this dilemma. The linearized
approximation was set up in such a way that $\Phi_6$ is close to its
background value $v$. It is precisely this assumption that breaks down near
$\Phi_6=0$, illustrated by a shaded region in figure \ref{figb}.

Presumably, the full non-linear solution to (\ref{eom}) will arrange
itself so that the $\Phi_6=0$ contour is coincident with the
horizon. Unfortunately, carrying out such an analysis is beyond our
immediate capabilities. One possible approach is to supplement the
expansion $\Phi_6 = v + \delta \Phi_6$ by a different expansion near
$\Phi=0$. Such an analysis indeed suggests that $\Phi_6=0$ is
coincident with the horizon. What is not trivial is to conclusively
argue that the solutions obtained by linearizing at different points
actually connect smoothly in the full solution. It would be
interesting to investigate these issues further.

With these disclaimers, however, we seem to be able to infer the basic
structure of the soliton solution from the embedding equation
(\ref{eom}). This can be viewed as a holographic confirmation of the
basic existence of these solitons established previously in
\cite{Cottrell:2014ura} for the case when $N=1$, as long as $c$ is finite.

The interesting tension with the conclusion of \cite{Mintun:2014aka},
however, has to do with understanding how these solitons behave in the
limit $c \rightarrow \infty$. If our soliton continues to exist in
that limit, we contradict the conclusions of \cite{Mintun:2014aka}. 

What we will show, in the next section, is that the embedding equation
(\ref{philin}) degenerates in the limit $ c \rightarrow \infty$. From
this, we conclude that 1) the system of \cite{Mintun:2014aka} is
singular without some UV completion, and that 2) while full string
dynamics can serve as a UV completion as was suggested in
\cite{Mintun:2014aka}, the integer tower of states (\ref{tower}) is
just as good as an alternate, economical, UV completion.

\section{The fate of the solitons in the $c \rightarrow \infty$ limit}

In this section, we will examine how the soliton solution is behaving
in the limit that $c$ is taken to be large, keeping $v$ and $\lambda$
fixed. It is useful to begin by looking closely again at
(\ref{philin}) and figure \ref{figb}. The large $c$ limit is
stretching the flat part of the graph in regions I and II.

We can zoom into the boundary of regions I and II by setting $v=0$ in
(\ref{philin}). Suppose we also separate variables and write
\be \delta \Phi_6 =  e^{i k x_3} \psi_k(\rho)\ .\ee
What one finds then is an equation of the form
\begin{equation}
\label{mathieu}
{1 \over \rho} \partial_\rho \left(\rho \partial_\rho  \psi_k\right)
-
k^2 \left(1+{1 \over c^2 \rho^4}\right)
 \psi_k =0
\end{equation}
which is actually Mathieu's equation \cite{Gubser:1998iu}. The $c
\rightarrow \infty$ limit then corresponds to the decoupling between
$\rho \ll 1/\sqrt{c}$ region and $\rho \gg 1/\sqrt{c}$ region precisely
in a manner analogous to how the asymptotically flat region and the
near horizon region decoupled in the original formulation of AdS/CFT
correspondence \cite{Maldacena:1997re}.

Let us examine this decoupling a little bit more closely. Let us define a parameter
\be \epsilon ={v^2 \over c} \ee
rescale to dimensionless coordinates
\be \rho = {v \over c} x, \qquad x_3 = {1 \over v} z \ee
and separate variables
\be \delta \Phi_6 = \int {dk \over 2 \pi} c_k e^{i k z} \psi_k(x) \ . \ee
Then, the equation for $\psi_k(x)$ reads
\begin{equation}
\label{philinResc}
\frac{1}{x}\partial_{x}\left(x \partial_{x}\psi_k\right)- k^2 \left(\epsilon^{2}+\frac{1}{\left(1+x^{2}\right)^{2}}\right)\psi_k=0 \ . 
\end{equation}

In these coordinates, for finite $\epsilon$, region I is $x \gg 1/\sqrt{\epsilon}$ whereas regions II and III are $x \ll 1/\sqrt{\epsilon}$. 

In region I, we can approximate the equation as
\begin{equation}
\frac{1}{x}\partial_{x}\left(x \partial_{x}\psi\right)-\epsilon^2 k^2 \psi=0
\end{equation}
which is solved by Bessel functions
\be \psi_k(x) = K_0(\epsilon k x)  \label{besselK} \ee
where we select the solution decaying at large $x$. 

In regions II and III, the equation truncates to 
\begin{equation}
\frac{1}{x}\partial_{x}\left(x \partial_{x}\psi_k\right)- \frac{k^2}{\left(1+x^{2}\right)^{2}}\psi_k=0
\end{equation}
which also admits a closed solution in terms of the Legendre functions
\be \psi_k(x) = \int {dk \over 2 \pi} \left( a_k P_{\lambda_k}\left(-1+{x^2 \over 1 + x^2}\right) +
b_k Q_{\lambda_k}\left(-1+{x^2 \over 1 + x^2}\right) \right) 
\ee 
with
\be \lambda_k = {1 \over 2}(-1+\sqrt{1-k^2}) \ . \ee

For small $\epsilon$, the solutions in regions I and the solutions in
regions II/III have overlapping regimes of validity in the region
\be k    < x < {1 \over \epsilon k} \label{ek} \ee
giving rise to a matching procedure, along the lines of
\cite{Klebanov:1997kc,Gubser:1998kv}, which relate $a_k$, $b_k$, and $c_k$. 

One can construct a reasonable approximation to the solution of
(\ref{philin}) by imposing a boundary condition for $\psi_k$ in
regions II/III so that it corresponds to a point-like source
near $x=0$ and to a decaying solution (\ref{besselK})
at $x \approx 1/\sqrt{\epsilon}$.  Carrying out this analysis
numerically to reproduce the features in figure \ref{figb} is rather
cumbersome, but one can identify basic features emerging in this type
of an analysis.

One sign that the $\epsilon \rightarrow 0$ limit is pathological can
be seen by noting that in the strict limit, flux in regions II/III can
no longer spread into region I. This can be confirmed in the solution
matching analysis outlined above, or can be thought of as the
consequence of decoupling. In this limit, flux in regions II/III can
only escape in the $x_3$ direction, giving rise to a linearly growing
potential
\be \delta \Phi_6= q v |z| \ . \ee
Such a solution is problematic in that 1) it does not asymptote to $v$
away from the source, and 2) has uniform energy density and can not
exist as a finite energy soliton. This is precisely the kind of
``confining'' behavior also encountered by \cite{Mintun:2014aka}.  In
other words, by taking a strict large $c$/small $\epsilon$ limit, we
have forced the solution, in region II/III, to change its asymptotic
behavior.  The absence of a solution with the prescribed asymptotic
behavior is a statement of the non-existence of the soliton solution.
From our point of view, we see the decoupling of the states (\ref{tower})
that provided the necessary ultra-violet completion is manifesting
itself as the decoupling of the soliton.  It is quite gratifying to
see similar pathologies arise from very different perspectives.

One can then interpret the small but finite $\epsilon$ as allowing the flux of the monopole to eventually escape into region I.  Once the flux escapes into region I, one can assess the ``effective charge distribution'' that can be inferred by probing the fields far in region I. That analysis leads to the conclusion that the charge source is smeared by scale set by
\be \Delta x^3 = {\sqrt{| \log \epsilon| } \over v} \ .  \ee
This can be confirmed, for example, by computing the width of the
$c_k$ distribution computed using the matching procedure outlined
above.  In particular, we find the $c_k$, expanded in $k$, has the form
\begin{equation}
c_{k}\sim -\frac{qv}{\pi} \left(1+\frac{1}{2}\ln(\epsilon)k^{2}+ {\cal O}(k^4)\right)
\end{equation}
for some very small $\epsilon$.  Such  logarithmic dependence in
$\epsilon$ arises because the expansion of the Bessel function
(\ref{besselK}),
\be  K_0(\epsilon k x) = 
 \left(-\log (x) -\gamma -\log \left(\frac{\epsilon  k}{2}\right)
    \right)+{\cal O}(x^1) \ . 
\ee

One can think of having a small $\epsilon$ suppressing the
penetration of the flux from region II/III into region I by a factor
of
\be {1 \over \sqrt{ | \log \epsilon | }} \ee
which is small in the strict $\epsilon \rightarrow 0$ limit, but only
logarithmically so. However, if we take the strict $\epsilon
\rightarrow 0$ limit while insisting on $\Phi_6 \rightarrow v$
asymptotic behavior, the soliton diffuses in the $x_3$ direction and
ceases to exist as a localized object.

This is precisely the sense in which the consistency between the
conclusion of \cite{Mintun:2014aka} and \cite{Cottrell:2014ura} is
maintained.

\section{Conclusion}
\label{conclusion}

In this and companion article \cite{Cottrell:2014ura}, we revisited
the issue of the decoupling of effective field theories on D3 and D3'
brane overlapping along 1+1 dimensions originally raised by
\cite{Mintun:2014aka}. In the original formulation, the angle between
the D3 and the D3'-brane was arranged to be perpendicular as outlined
in (\ref{orientation}). In such a setup, \cite{Mintun:2014aka} found
that it was not possible to consistently decouple stringy states and
obtain a closed dynamical system.

One main lesson from the work of \cite{Cottrell:2014ura} and this
article is that by scaling the angle between the D3 and the D3'-branes
(\ref{scaling}), one does arrive at a consistent decoupled system. The
difference between fixing the angle and fixing $a$ manifested in a
tower of 33' string states (\ref{tower}) surviving the $\alpha'
\rightarrow 0$ limit, so there are more states than were
envisioned in \cite{Mintun:2014aka}. These states appear to complete the UV dynamics so that the dynamics is closed
and compatible with S-duality. These are not so exotic either, having
been studied previously in various contexts
\cite{vanBaal:1984ar,Hashimoto:1997gm}.

In this article, we continued the work of \cite{Cottrell:2014ura} by
tracking how the soliton degenerates in the limit that $a \sim c
\rightarrow \infty$. This is the limit where all but the massless
state in the tower (\ref{tower}) become infinitely massive, and the
spectrum of light states approaches that which was considered in
\cite{Mintun:2014aka}.  In order to make this analysis tractable, we
generalized to the setup (\ref{orient2}) where the number $N$ of D3s is
taken to be large so that it can be described holographically, and
studied the D3'-brane as a probe being embedded into $AdS_5 \times S^5$.
What we found is that in the large $c$ limit, the soliton delocalizes
and disappears as a state.

As an exercise in holographic embedding, one could have just as easily
started with the 90 degree embedding (\ref{orientation}) and reached a
similar conclusion.  Consider, first, the supergravity solution for the full D3-brane prior to taking the near horizon limit
\be ds^2= f^{-1/2} (-dt^2 + d\vec x^2) + f^{1/2} (dr^2 + d \Omega_5^2), \qquad f = 1 +{R^4 \over r^4} \ee
and embed a D3 oriented on $t$, $x_3$, $x_4$, and $x_5$. Using polar
coordinates $(\rho,\varphi)$ to parameterize the $(x_4,x_5)$ plane, the 
embedding equation takes the form
\begin{equation}
\frac{1}{{\rho}}\partial_{{\rho}}\left({\rho}\partial_{{\rho}}\Phi_{6}\right)+\partial_{3}\left(1+\frac{R^{4}}{\left({\rho}^2+R^{4}\Phi_{6}^{2}\right)^{2}}\right)\partial_{3}\Phi_{6}=0 \ . 
\end{equation}
This equation is identical to (\ref{eom}) upon substituting $c
=1/R^2$. The act of taking the $\alpha' \sim R^2 \rightarrow 0$ is
having the same effect of decoupling the near horizon and the
asymptotic region as was seen in the $c \rightarrow \infty$ limit, and
in the process, a soliton that would have existed for finite $R^2$ is
also decoupling. The disadvantage of working with a fixed angle like
this, however, is that one misses the possibility of finding the tower of
states (\ref{tower}) as an economic alternative to invoking string
dynamics in regulating the dynamics.

In this article, we mainly focused on the magnetic soliton, but it is
just as straightforward to analyze DBI embedding corresponding to an
electric source. The supersymmetry and the resulting constraint will
take on slightly different form, but we arrive at the same embedding
equation (\ref{eom}). The only difference for the electric case as
opposed to the magnetic case is the normalization of charge. Instead
of (\ref{quantcond}), we find
\be q = {\pi k g_s  \over \sqrt{\lambda}}  \ . \ee
Strictly speaking, in the 't Hooft limit where $N \rightarrow \infty$
keeping $\lambda$ fixed, this approaches zero. This is simply the
reflection of the fact that when $g_s$ is small, the fundamental string is
less tense than a D-string. When considering the case were $N$ is
large but finite, however, there will be some bending of the D3'-brane 
due to the tension of the fundamental string. Our analysis then
implies that states with electric charges, i.e. the $B$ and the $C$
fields, are delocalizing and decoupling in the $c\rightarrow
\infty$ limit, and in the setup of \cite{Mintun:2014aka}. This is not
surprising in light of the fact that the system under consideration is
S-dual. If the magnetic state is decoupling, so must its electric
dual.

Ultimately, we are finding that the effective field theory in the $c
\rightarrow \infty$ limit in terms of $U(N)\times U(1)$ fields and
bifundamentals ceases to exist because both the magnetic duals of the
bifundamental fields, and the bifundamental fields themselves,
decouple from the spectrum. That does not mean that an effective
dynamical description do not exist.  After all, an embedding of D3' in
$AdS_5 \times S^5$ does exist, and there are some effective dynamics for
the 3'3' strings which couples non-trivially to bulk states in $AdS_5
\times S^5$. All that we have shown is that the system does not admit
33' state in the electric or the magnetic sector. It should be noted,
however, that in such a holographic setup, the asymptotic behavior of
open strings ending on the D3' is modified drastically, in the sense
that the D3' brane hits the boundary of $AdS_5\times S^5$. It is
possible that a similar change in the asymptotic behavior should be
expected in the $U(N)$ sector when $N$ is of order one. It is not
clear how one would describe such a system in terms of field
theory. One possible scenario is that the effective dimension of the
D3'-brane dynamics becomes 1+1 dimensional in light of being confined
inside a finite $AdS$ box, and as such, experience strong quantum
fluctuation and flow essentially to a D1-D5 conformal field theory in
the IR. It would be very interesteing to understanding this issue
better.

One of the most notable features of the system oriented as
(\ref{orient1}) and scaled as (\ref{scaling}) is the tower of states
(\ref{tower}) which appears to be playing an indispensable role in
regulating the UV dynamics. This tower can also be thought of as a
{\it single} Regge trajectory. Unlike string theory which regulates
the UV dynamics (say, of gravity) with an infinite set of Regge
trajectories, here we achieve regularity with a single trajectory. To
the best of our knowledge, this is the first time such a
regularization has been seen at work. The signal that the theory is sick
was extracted from subtle non-perturbative features encoded in soliton
dynamics. It would be interesting to find a perturbative manifestation
of these singularities and to better understand how a tower like
(\ref{tower}) is regulating it. Perhaps one can find some hints by
analyzing charge renormalization or photon vacuum polarization for the
electric theory. We hope to address these questions in the near
future.

\section*{Acknowledgements}

We thank 
M.~Kruczenski,
P.~Ouyang,
M.~Yamazaki
for discussions. This work was supported in part by funds from the University of Wisconsin, Madison.

\appendix
\section{Supersymmetry Conditions}
\label{appendixA}

In this Appendix we will show more explicitly how the BPS conditions (\ref{finalansatz}) are derived.  The basic strategy is as follows.  For a supersymmetric D-brane embedding, the supersymmetry generator, $\epsilon$ must obey a $\kappa$ projection condition, written as:
\begin{equation}
\Gamma(\Phi_{i},F)\epsilon = \epsilon \ . 
\end{equation}
The supersymmetry generator $\epsilon$ must also be a generator for
the supersymmetry of the ambient $AdS$ space localized to the brane.
If we are given a brane embedding specified by $\Phi_{i}$ and
$F_{\mu\nu}$ then the $\kappa$ projector may be used to determine
which spinors are supersymmetry generators.  On the other hand, if we
know the supersymmetry generators in advance, then the equation
$\Gamma(\Phi_{i},F) \epsilon = \epsilon$ may be viewed as a constraint
on the fields $\Phi_{i}$ and $F_{\mu\nu}$.  Since $\epsilon$ is $16$
dimensional, in principle we have $16$ (complex) constraints for each
preserved supersymmetry.  Our goal then, is to first determine which
supersymmetries should be preserved by the tilted monopole solution
and then to view the $\kappa$ projection condition as a set of
equations to solve for the embedding coordinates and fluxes.  On
general grounds, one expects that the allowed generators for the
tilted monopole lie in the intersection of the supersymmetry
generators for a pure tilted brane (no monopole) and a pure monopole
(no tilt).  We must therefore analyze these two cases separately and
look for supersymmetry generators preserved by both.

We begin by reviewing kappa symmetry in $AdS_{5}$, following \cite{Skenderis:2002vf}.  The basic definition of the kappa projector, $\Gamma$, is:
\begin{equation}
\label{kappa}
d^{4}\xi \Gamma = - e^{-\Phi} \mathcal{L}_{DBI}^{-1} e^{\mathcal{F}}\wedge X|_{vol}
\end{equation}
with 
\begin{eqnarray}
\label{xgammadef}
X&\equiv& \oplus_{n}\Gamma_{(2n)}K^{n} I \\ \nonumber
\Gamma_{(n)}&\equiv&\frac{1}{n!}d\xi^{i_{n}}\wedge ...\wedge d\xi^{i_{1}} \Gamma_{i_{1}...i_{n}}
\end{eqnarray}
where $K$ acts as complex conjugation and $I\psi=-i\psi$.  Unless stated otherwise, we will use the following basis for gamma matrices
\begin{eqnarray}
\Gamma_{\mu}&=&R u \gamma_{\mu}\qquad \qquad \mu=0\ldots3 \nonumber \\
\Gamma_{\rho}&=&R u \gamma_{1} \nonumber \\
\Gamma_{\varphi}&=& R u \rho \gamma_{2} \nonumber \\
\Gamma_{u}&=&  R u^{-1} \gamma_{4} \nonumber \\
\Gamma_{\phi}&=& R \sin\theta\gamma_{5} \nonumber \\
\Gamma_{\theta}&=&R \gamma_{6} \ . 
\end{eqnarray}
The condition for a supersymmetric embedding is that there exist a Weyl spinor $\epsilon$ such that $\Gamma \epsilon = \epsilon$ and where $\epsilon$ satisfies the background Killing spinor equations in $AdS^{5}\times S^{5}$.  
\begin{equation}
\label{killingeq}
\left(D_{M}+\frac{i}{2}\gamma^{01234}\Gamma_{M}\right)\epsilon=0 \ . 
\end{equation}
A full set of background supersymmetry generators satisfying (\ref{killingeq}) was found in \cite{Claus:1998yw}.  Following the notation of \cite{Skenderis:2002vf} we may write the solutions as
\begin{eqnarray}
\epsilon_{+}&=&-u^{-1/2}\gamma_{4} h(\theta) \eta_{2} \\ \nonumber
\epsilon_{-}&=& u^{1/2} h(\theta)\left(\eta_{1}+x \cdot \gamma \eta_{2}\right)  \nonumber
\end{eqnarray}
where $\eta_{1}$, $\eta_{2}$ are spinors of negative and positive chirality underr the $(3+1)d$ chirality operator.   A complete expression for $h(\theta_{i})$ was provided in \cite{Skenderis:2002vf}.  Here, we have truncated to $x^{7,8,9}=0$ and will only need the following formula:
\begin{equation}
\label{heq}
h(\theta_{i})=e^{\frac{1}{2}\theta \gamma_{46}}e^{-\frac{1}{2}\phi \gamma_{56}} \
 . 
\end{equation}
It will be useful to further decompose $\eta_{i}$ into real spinors, $\lambda$, $\eta$ as follows:
\begin{eqnarray}
\eta_{1}&=&\lambda-i \gamma^{0123} \lambda \\ \nonumber
\eta_{2}&=&\eta+i \gamma^{0123}\eta \ . 
\end{eqnarray}
The generator parameterized by $\eta$ corresponds to superconformal generators and these are generically broken for the solutions we are interested in.   We will thus only consider supersymmetries generated by $\lambda$ in what follows.  

The kappa projector in its general form (\ref{kappa}) suffers from the
drawback that it depends upon $\mathcal{L}_{DBI}$, which cannot be
written in a usable form without knowing the solution ahead of time.
To work around this, we will derive a simplified form of (\ref{kappa})
that does not depend on $\mathcal{L}_{DBI}$.  Although we will
apparently lose some of the information while doing this, we will find
that our simplified condition still has enough constraints to
determine the BPS conditions uniquely.

We start by expanding (\ref{kappa})
\begin{eqnarray}
\Gamma&=&-\mathcal{L}_{DBI}^{-1}\left(\Gamma_{(4)}K^{2}I+2\pi \alpha' F_{2}\wedge \Gamma_{(2)} K I\right) \ . 
\end{eqnarray}
When acting on $\epsilon$ we may decompose into real and imaginary parts and then subtract the two.  The real equation is
\be
\label{reeq}
\mathcal{L}_{DBI}^{-1}\left(\Gamma_{(4)}+2\pi \alpha' F_{2}\wedge \Gamma_{(2)}\right)\left(h\gamma^{0123}\lambda\right)=h\lambda
\ee
and the imaginary equation is:
\be
\label{imeq}
\mathcal{L}_{DBI}^{-1}\left(\Gamma_{(4)}-2\pi\alpha' F_{2}\wedge \Gamma_{(2)}\right)\left(h\lambda\right) =-h\gamma^{0123}\lambda \ . 
\ee
Now we multiply both sides of the second equation above, (\ref{imeq}),
by $\gamma^{0123}$ remembering that $(\gamma^{0123})^{2}=-1$ and that
we must anticommute past the $x\cdot\vec{\gamma}$.  Let us also define
\begin{eqnarray}
\widetilde{\Gamma}_{4}&=&-\gamma^{0123}\Gamma_{4}\gamma^{0123} \\ \nonumber
\widetilde{\Gamma_{2}\wedge F_{2}}&=&- \gamma^{0123}\Gamma_{2}\wedge F_{2} \gamma^{0123} \ . 
\end{eqnarray}
Equation (\ref{imeq}) then becomes
\begin{eqnarray}
\label{newimeq}
\mathcal{L}_{DBI}^{-1}\left(\widetilde{\Gamma}_{(4)}-2\pi \alpha'\widetilde{F_{2}\wedge \Gamma_{(2)}}\right)\left(h\gamma^{0123}\lambda\right) &=&h\lambda   \ . 
\end{eqnarray}
Now we subtract equation (\ref{newimeq}) from (\ref{reeq}).  This will give us our main equation 
\begin{equation}
\label{master}
\left(\left(\Gamma_{(4)}-\widetilde{\Gamma}_{(4)}\right)+2\pi\alpha'\left(F_{2}\wedge\Gamma_{(2)}+\widetilde{F_{2}\wedge\Gamma_{(2)}}\right)\right)
h\gamma^{0123}\lambda=0 \ . 
\end{equation}
We have thus succeeded in eliminating $\mathcal{L}_{DBI}$.  The task
is now to determine which $\lambda$'s must be annihilated by the
operator appearing above. Once we know this, then (\ref{master}) may
be looked at as a set of algebraic constraints on the various fields
living inside of $\Gamma_{(4)}$, $\Gamma_{(2)}$ and $F_{2}$.  These
will be our BPS conditions.  Note that since, in principle, the
equation above only contains half the constraints of the original
kappa projector we will be obliged to check that our final solutions
still satisfy the full set of constraints.  Indeed, we will find that
this holds in all cases.  We must now determine the appropriate set of
$\lambda$'s by looking at the monopoles and tilted brane separately:

\subsection{Magnetic monopole on untilted brane}
\label{monopolesubsection}

In the case of a single monopole with no tilt one expects that
$\theta=0$ and that the solution is a function of (world volume)
radius only.  The only non-trivial variables is $u$, which in this
case is equal to $\Phi_{6}$.  The kappa symmetry analysis gives the
BPS conditions
\begin{equation}
\label{schwarzkappa}
\left(R^{2}\sin(\theta')(r')^{2}\partial_{r'}u \,\gamma_{1234}+2\pi\alpha' F_{\theta'\varphi}\right)h\gamma^{0123}\lambda=0 \ .
\end{equation}
Since the eigenvalues of $\gamma_{1234}$ are $\pm1$, the above equation can only have a solution if:
\begin{eqnarray}
\label{schwarzbps}
\gamma_{1234}h\gamma^{0123}\lambda&=&\pm h\gamma^{0123}\lambda \\ \nonumber
2\pi\alpha' F_{\theta'\varphi}&=& \mp R^{2}\sin \theta' (r')^{2}\partial_{r'}u \ . 
\end{eqnarray}
Without loss of generality, we will choose the upper sign.  The
condition on $\lambda$ may be rewritten as
\begin{equation}
h^{-1}\gamma_{1234}h\lambda=\lambda \ . 
\end{equation}
Plugging $\theta=0$ into expression (\ref{heq}) (as is appropriate for
a brane with no tilt) one finds that this is equivalent to:
\begin{equation}
\label{susymonopole}
\gamma_{1234}\lambda=\lambda \ . 
\end{equation}
This is the final condition that we will need. 

\subsection{Tilted Brane}
\label{tiltedbranesubsection}

We now repeat the procedure above for tilted branes with no monopole.  Assuming $F=0$ and $\theta=\pi/2$ one finds the BPS conditions: 
\begin{equation}
\left(\rho u' \gamma_{42}-u\gamma_{15}\right) h \gamma^{0123}\lambda=0 \ . 
\end{equation} 
Again, relying on the fact that $\gamma_{1245}$ has eigenvalues of
$\pm 1$, this leads to the following BPS conditions:
\begin{eqnarray}
\label{tiltbps}
\gamma_{1245}h\gamma^{0123}\lambda&=&\pm h\gamma^{0123}\lambda \\ \nonumber
\rho u'&=&\pm u \ . 
\end{eqnarray}
In this note, we will be interested in the $+$ sign solution, which
leads to tilted brane solutions of the form $u=c\rho$.  (The minus
sign embeding of the form $u=c\rho^{-1}$ which was interpreted as
surface operators in in \cite{Drukker:2008wr}.)  Recalling also that
$u=\sqrt{(x^{4})^{2}+(x^{5})^{2}}/R^{2}$ we can get an expression for
the tilt angle as:
\begin{equation}
\tan\gamma=R^{2}c \ . 
\end{equation}
Finally, we need to determine the appropriate condition on $\lambda$.  Using the formula (\ref{heq}) when $\theta=\pi/2$ we find that the first condition in (\ref{tiltbps}) becomes
\begin{equation}
\label{tiltsusycond}
\gamma_{1256}\lambda=-\lambda \ .
\end{equation}

\subsection{Magnetic monopole on tilted brane}

For this case, we impose conditions 
(\ref{susymonopole}) and (\ref{tiltsusycond}) simultaneously. This applied to (\ref{master}) will give rise to the supersymmetry constraint (\ref{finalansatz}). 

\subsection{Electric Monopole}
\label{electricsubsection}

With minor  modification, the above analysis can be extended to the case of an electric monopole and show that it has the properties consistent with S-duality.
The general spinor condition (\ref{tiltbps}) from the tilting of the D3' is the same as the magnetic monopole case. The presence of an electric charge, on the other hand, gives rise to a constraint
\begin{eqnarray}
\label{electricsusycond}
\gamma_{04}\lambda&=&\lambda \\ \nonumber
2\pi\alpha'F_{0r'}&=&R^2 \partial_{r'}u \ . 
\end{eqnarray}
Requiring that spinors satisfy both (\ref{tiltsusycond}) and
(\ref{electricsusycond}) and then reading off the supersymmetry
condition from (\ref{master}) gives the electric analogue of
(\ref{finalansatz}),
\begin{eqnarray}
2\pi\alpha'F_{0\rho}&=&R^2(\partial_\rho u-\frac{u}{\rho}\sin^2\theta)\sec\theta \nonumber \\
2\pi\alpha'F_{03}&=&R^2\partial_3 u\sec\theta \nonumber \ . 
\end{eqnarray}
As in the magnetic case, the $\theta$ equations imply that $u \sin\theta=c\rho$, and therefore that the equations above may be written as 
\begin{eqnarray}
\label{electricbps}
F_{0\rho}&=&\frac{R^2}{2\pi\alpha'}\partial_\rho\Phi_6 \\ \nonumber
F_{03}&=&\frac{R^2}{2\pi\alpha'}\partial_3\Phi_6 \ . 
\end{eqnarray}
Now it is clear that the Bianchi identity $dF=0$ will be trivial.  One
may also check that the equation of motion for $F$ obtained by varying
the full action will give precisely the equation of motion derived
previously in the magnetic case, i.e., (\ref{eom}).  However, as was
noted previously, one cannot obtain the full equations of motion by
varying the action obtained after substituting in the BPS ansatz.  In
the electric case, substituting the ansatz back into (\ref{action})
rise to the following trivial Lagrangian,
\begin{equation}
\mathcal{L} = -\frac{\lambda}{8\pi^{3}g_{s}} \rho c^{2} \ , 
\end{equation}
but the Hamiltonian has the form (\ref{bpsaction}) identical to the
one encountered earlier in the magnetic case.

\bibliography{soliton2}\bibliographystyle{utphys}

\end{document}